\documentclass{aa}  

\usepackage{natbib}
\usepackage{graphicx}
\usepackage{amssymb}
\usepackage{txfonts}
\newcommand{\msun}{M$_{\odot}$\,}
\newcommand{\mstar}{M$_{\star}$}
\newcommand{\rstar}{R$_{\star}$}
\newcommand{\rsun}{R$_{\odot}$}

\newcommand{\ms}{m\,s$^{-1}$\,}
\newcommand{\teff}{T$_{{\rm eff}}$\,}
\newcommand{\logg}{\ensuremath{\log{g}}}
\newcommand{\feh}{\ensuremath{{\rm [Fe/H]}}}
\newcommand{\zaspe}{{\small\texttt{ZASPE}}}
\newcommand{\moog}{{\small\texttt{MOOG}}}
\newcommand{\ceres}{{\small\texttt{CERES}}}
\newcommand{\exonailer}{{\small \texttt{EXONAILER}}}
\newcommand{\radvel}{{\small\texttt{RAD-VEL}}}
\newcommand{\batman}{{\small\texttt{BATMAN}}}
\newcommand{\gaia}{{\small\texttt{Gaia}}} 
\newcommand{\vsini}{\ensuremath{v \sin{i}}}

\newcommand{\stnameA}{HD\,2685}

\newcommand{\plnameA}{HD~2685~$b$}

\renewcommand{\cite}{\citealp}

\begin{document} 

   \title{\plnameA\,: A Hot-Jupiter orbiting an early F-type star detected by TESS\thanks{Tables of photometry are available in the electronic version of this paper.}}

    \author{Mat\'ias I. Jones \inst{1} 
    \and Rafael Brahm \inst{2,3}    
    \and Nestor Espinoza \inst{4}  
    \and Songhu Wang  \inst{5}\thanks{51 Pegasi b Fellow}     
    \and Avi Shporer \inst{6} 
    \and Thomas Henning \inst{4} 
    \and Andr\'es Jord\'an \inst{2,3} 
    \and Paula Sarkis \inst{4} 
    \and Leonardo A. Paredes \inst{7} 
    \and James Hodari-Sadiki  \inst{7} 
    \and Todd Henry  \inst{8} 
    \and Bryndis Cruz \inst{5} 
    \and Louise D. Nielsen \inst{9} 
    \and Fran\c cois Bouchy \inst{9} 
    \and Francesco Pepe \inst{9} 
    \and Damien S{\'e}gransan \inst{9} 
    \and Oliver Turner \inst{9} 
    \and St{\'e}phane Udry \inst{9} 
    \and Gaspar Bakos \inst{10} 
    \and David Osip \inst{11} 
    \and Vincent Suc \inst{2} 
    \and Carl Ziegler \inst{12} 
    \and Andrei Tokovinin \inst{13} 
    \and Nick M. Law \inst{14} 
    \and Andrew W. Mann \inst{14} 
    \and Howard Relles \inst{6} 
    \and Karren A. Collins \inst{15} 
    \and Daniel Bayliss \inst{16} 
    \and Elyar Sedaghati \inst{1} 
    \and David W.~Latham \inst{15} 
    \and Sara Seager \inst{6,17} 
    \and Joshua N.~Winn \inst{18} 
    \and Jon M.~Jenkins \inst{19} 
    \and Jeffrey C.~Smith \inst{19,20} 
    \and Misty Davies \inst{19} 
    \and Peter Tenenbaum \inst{19,20} 
    \and Jason Dittmann\inst{6}**  
    \and Andrew Vanderburg\inst{21}\thanks{NASA Sagan Fellow} 
    \and Jessie L. Christiansen\inst{22} 
    \and Kari Haworth\inst{6} 
    \and John Doty\inst{6} 
    \and Gabor Furesz\inst{6} 
    \and Greg Laughlin\inst{5} 
}

\institute{European Southern Observatory, Alonso de C\'ordova 3107, Vitacura, Casilla 19001, Santiago, Chile \\\email{mjones@eso.org}
\and Instituto de Astrof\'isica, Facultad de F\'isica, Pontificia Universidad Cat\'olica de Chile, Av. Vicu\~na Mackenna 4860, 7820436 Macul, Santiago, Chile
\and Millennium Institute of Astrophysics, 7820436 Santiago, Chile
\and Max-Planck-Institut fur Astronomie, K{\"o}nigstuhl 17, D-69117 Heidelberg, Germany
\and Department of Astronomy, Yale University, New Haven, CT 06511, USA
\and Department of Physics and Kavli Institute for Astrophysics and Space Research, Massachusetts Institute of Technology, Cambridge, MA
02139, USA   
\and Physics and Astronomy Department, Georgia State University, Atlanta, GA 30302, USA
\and RECONS Institute, Chambersburg, PA, USA
\and Geneva Observatory, University of Geneva, Chemin des Mailettes 51, 1290 Versoix, Switzerland
\and Department of Astrophysical Sciences, Princeton University, NJ 08544, USA   
\and Las Campanas Observatory, Carnegie Institution of Washington, Colina el Pino, Casilla 601 La Serena, Chile   
\and Dunlap Institute for Astronomy and Astrophysics, University of Toronto, Ontario M5S 3H4, Canada 
\and Cerro Tololo Inter-American Observatory, Casilla 603, La Serena, Chile
\and Department of Physics and Astronomy, University of North Carolina at Chapel Hill, Chapel Hill, NC 27599-3255, USA
\and Harvard-Smithsonian Center for Astrophysics, 60 Garden Street, Cambridge, MA 02138 USA
\and Department of Physics, University of Warwick, Gibbet Hill Rd., Coventry, CV4 7AL, UK
\and Department of Earth and Planetary Sciences, MIT, 77 Massachusetts Avenue, Cambridge, MA 02139, USA
\and Department of Astrophysical Sciences, Princeton University, 4 Ivy Lane, Princeton, NJ 08544, USA
\and NASA Ames Research Center, Moffett Field, CA 94035, USA
\and SETI Institute, 189 Bernardo Avenue, Suite 100, Mountain View, CA 94043, USA
\and Department of Astronomy, The University of Texas at Austin, Austin, TX 78712, USA
\and IPAC, Mail Code 100-22, Caltech, 1200 E. California Blvd. Pasadena, CA 91125, USA
}
   \date{}

 
  \abstract{We report on the confirmation of a transiting giant planet around the relatively hot (\teff = 6801 $\pm$ 56\,K) star \stnameA\,, whose transit signal was detected in Sector 1 data of the TESS mission. 
  We confirmed the planetary nature of the transit signal by using Doppler velocimetric measurements with CHIRON, CORALIE and FEROS, as well as photometric data with CHAT and LCOGT. From the photometry and radial velocities joint analysis, we derived the following parameters for \plnameA: $P$=4.12692$\pm$0.00004 days, M$_P$=1.18 $\pm$ 0.09 $M_J$ and $R_P$=1.44 $\pm$ 0.01 $R_J$. 
  This system is a typical example of an inflated transiting Hot-Jupiter in a circular orbit. Given the host star apparent visual magnitude ($V$ = 9.6 mag), this is one of the brightest known stars hosting a transiting {\it Hot-Jupiter}, and a good example of the upcoming systems that will be detected by TESS during the two-year primary mission. 
  This is also an excellent target for future ground and space based atmospheric characterization as well as a good candidate for measuring the projected spin-orbit misalignment angle via the Rossiter-McLaughlin effect.  }


    \maketitle
%

\section{Introduction}
\label{sec:intr}

Transiting planets are of great importance because they provide us with crucial information about their formation and evolution. When transit observations are complemented with precision radial velocity (RV) data, it is possible to accurately determine their mass and radius, and therefore their mean density. Using this information we can study their internal structure, and compare their parameters with those predicted by theoretical planetary structure models (\citealt{baraffe2014}; \citealt{thorngren2016}). In addition, due to their proximity to the parent star, we can study how they are affected by the strong stellar irradiation (e.g. \citealt{guillot2002}) and tidal interactions with the host star (\citealt{rasio1996}; \citealt{matsumura2010}). Moreover, by measuring the stellar obliquity, we can
further study different migration scenarios (for a review see \citealt{winn2015}).
\newline \indent
During the past decade, the transit method has become the most efficient way to detect compact planetary systems. While it is still challenging to detect transit events from the ground, particularly for sub-mmag transit depths and/or long-duration events, the advent of dedicated space missions has revolutionized the way we study close-in extrasolar planets. 
In particular, the $Kepler$ mission \citep{kepler} launched in 2009, provided us with thousands of transiting systems\footnote{https://keplerscience.arc.nasa.gov/}, including Earth-like, Neptune-sized, and gas giant planets in short-period orbits, also revealing that the majority of the stars in our Galaxy host planets, and in particular the M-dwarfs (\citealt{dressing2013}; \citealt{mulders2015}).
Unfortunately, most of the candidate systems detected by $Kepler$ and its mission extension $K2$ are relatively faint, thus their detailed  characterization via radial velocity follow-up and transmission spectroscopy is restricted only to the brightest examples. \newline \indent
On the other hand, NASA's Transiting Exoplanet Survey Satellite (TESS; \citealt{tess}), launched in 2018 and already in full operations, will target more than 200,000 stars at 2-minute cadence, 5\% of them brighter than $\sim$ 8 mag.
Among these bright stars, a total of $\sim$ 100 planets are expected to be detected, with $\sim$ 7 of them orbiting stars with $V \lesssim$ 6 mag (\citealt{barclay2018}).
The recently published inner planet orbiting the naked-eye star $\pi$ Mensae, detected in TESS's Sector 1, is a good example of  this (\citealt{gandolfi2018,huang2018}). 
These newly detected bright transiting systems will be primary targets for further transit studies and detailed atmospheric characterization by ground-based high-resolution transmission spectroscopy and future space missions
such as the CHaracterising ExOPlanet Satellite (CHEOPS; \citealt{cheops}) and the James Web Space Telescope (JWST; \citealt{jwst}). These systems will also provide us with the opportunity of a dedicated spectroscopic follow-up from the ground, 
to measure their individual masses (particularly for Earth-sized planets) and to study the host star spin-orbit alignment (obliquity) via the Rossiter-McLaughlin effect (e.g. \citealt{queloz2000}). \newline \indent
In this paper we report on the spectroscopic confirmation of a transiting {\it Hot-Jupiter} around the early F-type star \stnameA\,(TIC\,267263253, TOI\,135), discovered in Sector 1 data of the TESS mission. After HD\,202772\,A\,b (\citealt{wang2018}), this is the second {\it Hot-Jupiter} detected by the TESS mission. With a visual magnitude V = 9.6, \stnameA\, is also among the brightest known stars that host a {\it Hot-Jupiter}, making this system an ideal target for a detailed follow-up characterization, and a good example of the upcoming transiting planets that will be detected by TESS. \newline \indent
The paper is structured as follows. In Section 2 we describe the photometric and spectroscopic observations as well as the data analysis. In Section 3 we derive the stellar properties. In Section 4 we present the global modelling using the combined radial velocities and transit photometry. 
Finally, the discussion is presented in Section 5.

\section{Observations and data analysis}
\label{sec:obs}

\begin{figure*}[h!]
\vspace{0cm}\hspace{0cm}
\includegraphics[scale=0.50,angle=0]{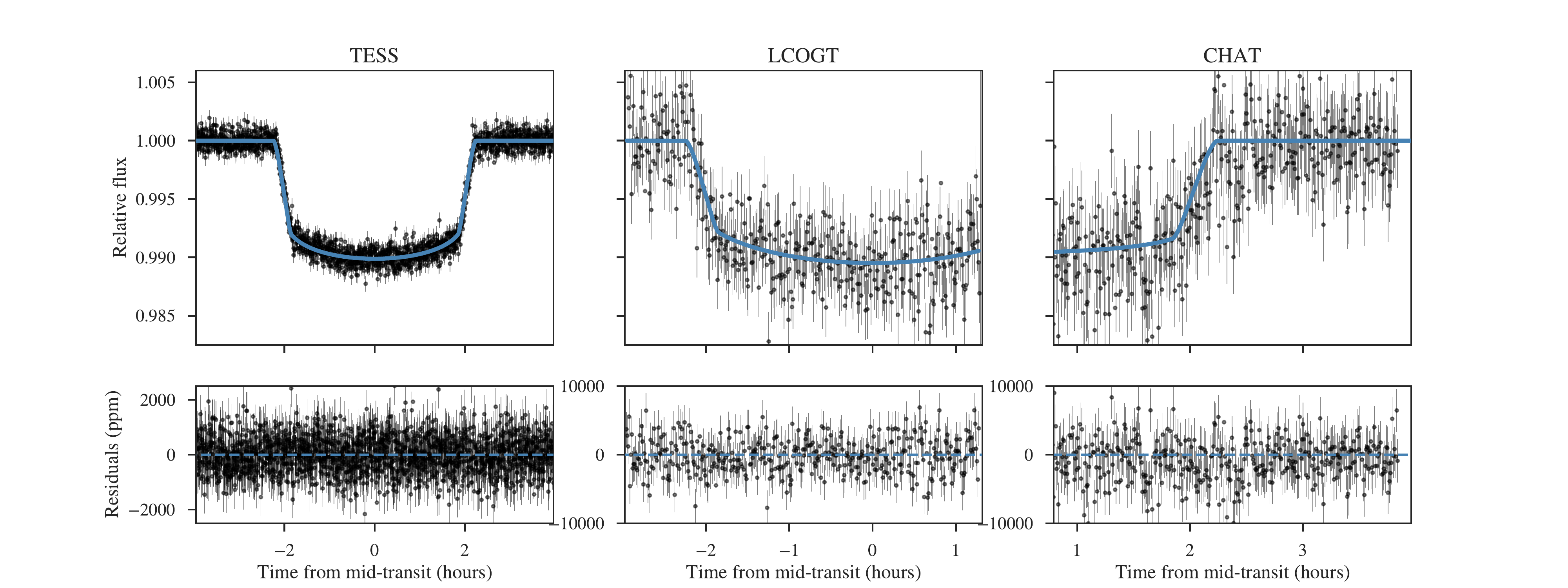}
\caption{{\it Left:} Detrended and phase-folded TESS light curve of \stnameA\, during the transit. The solid blue line corresponds to the transit fit from the joint analysis. 
{\it Middle:} LCOGT photometric data covering the ingress and mid part of one individual transit. {\it Right} CHAT data covering the egress of the same transit event observed by LCOGT. 
\label{fig:tess1}}
\end{figure*}

\subsection{TESS photometric data}

\stnameA\, (TIC\,267263253, TOI\,135) was observed in the high-cadence mode (2-minute exposures) in Sector 1 of the TESS space mission. These observations were collected by Camera 3, between July 25 and August 22, 2018. \stnameA\ is not planned to be observed in other TESS sectors during the TESS primary mission. The light curve was processed by the Science Processing Operations Center (SPOC) pipeline \citep{jenkins2016}, which is based on the Kepler Mission science pipeline (\citealt{jenkins2017}) and made available by the NASA Ames SPOC center and at the MAST archive\footnote{https://archive.stsci.edu/}. 
This dataset is comprised of a total of 18097 individual measurements, in which a total of 7 transit events were identified, with depths of $\sim$ 10,000 ppm and  duration of $\sim$ 4.5 hours. After masking out the transits, we detrended the light curve using a gaussian process modelling, with the quasi-periodic kernel presented in \citep{foreman-mackey2017}.
Figure \ref{fig:tess1} shows the detrended and phase-folded TESS light curve of \stnameA. The blue line corresponds to the best-fitting model described in section \ref{sec:joint_fit}. 

\subsection{CHAT photometry}

In addition to the TESS photometric data, we also collected a total of 411 ground-based measurements using the Chilean-Hungarian Automated Telescope (CHAT; Jord\'an et al. 2018, in prep.), installed at Las Campanas Observatory, in Chile. Observations were acquired on September 22 UT of 2018 using the i' sloan filter. The exposure time per image was of 14 s, which translated in a cadence of $\approx$25 s. A mild defocus was applied. The data were reduced through a dedicated  automatic pipeline \citep{hartman:2018,jordan:2018}. 
Thanks to the spatial resolution of CHAT (pixel scale of 0.6 arcsec), these were the first observations that helped us to confirm the source of the transit signal, discarding other potential sources in the TESS field of view. 
Figure \ref{fig:tess1} shows the CHAT photometry of \stnameA\, during the transit. The residual scatter is 3105 ppm and 1001 ppm when binned to 5 minutes bins. As can be seen, even though the CHAT data only covered the second half of one transit event, the observed transit depth is consistent with the TESS data.

\subsection{Las Cumbres Observatory photometry}

We performed ground-based photometric follow-up using the Las Cumbres Observatory Global Telescope (LCOGT\footnote{http:lco.gobal}) network \citep{brown2013}. On 2018 September 22 UT, we observed an almost complete transit (missing the egress) in the $i$-band using the LCOGT 0.4-m telescopes situated at South Africa Astronomical Observatory (SAAO) at Sutherland, South Africa. This observations were taken the same night as the CHAT observations, meaning that we covered one full transit using two different instruments.
The observations were done with the SBIG camera and consisted of 394 exposures with an exposure time of 30s, taken while the telescope was defocused by 1.5mm to spread the stellar point-spread function over more pixels, thereby reducing the impact of flat-fielding uncertainties. The data was reduced by the LCOGT pipeline and photometry was carried out with AstroImageJ \citep{collins2017}. The light curve is plotted in Fig.~\ref{fig:tess1}.
The residual scatter is 2560 ppm per point and 786 ppm when binned to 5 minutes bins. These observations were made as part of an LCOGT Key Project to follow-up TESS transiting planet candidates and characterize transiting planets using the LCOGT network\footnote{{space.mit.edu/$\sim$shporer/LCOKP}}. 

\subsection{Speckle imaging}

\begin{figure}
\vspace{0cm}\hspace{0cm}
\includegraphics[scale=0.6,angle=0]{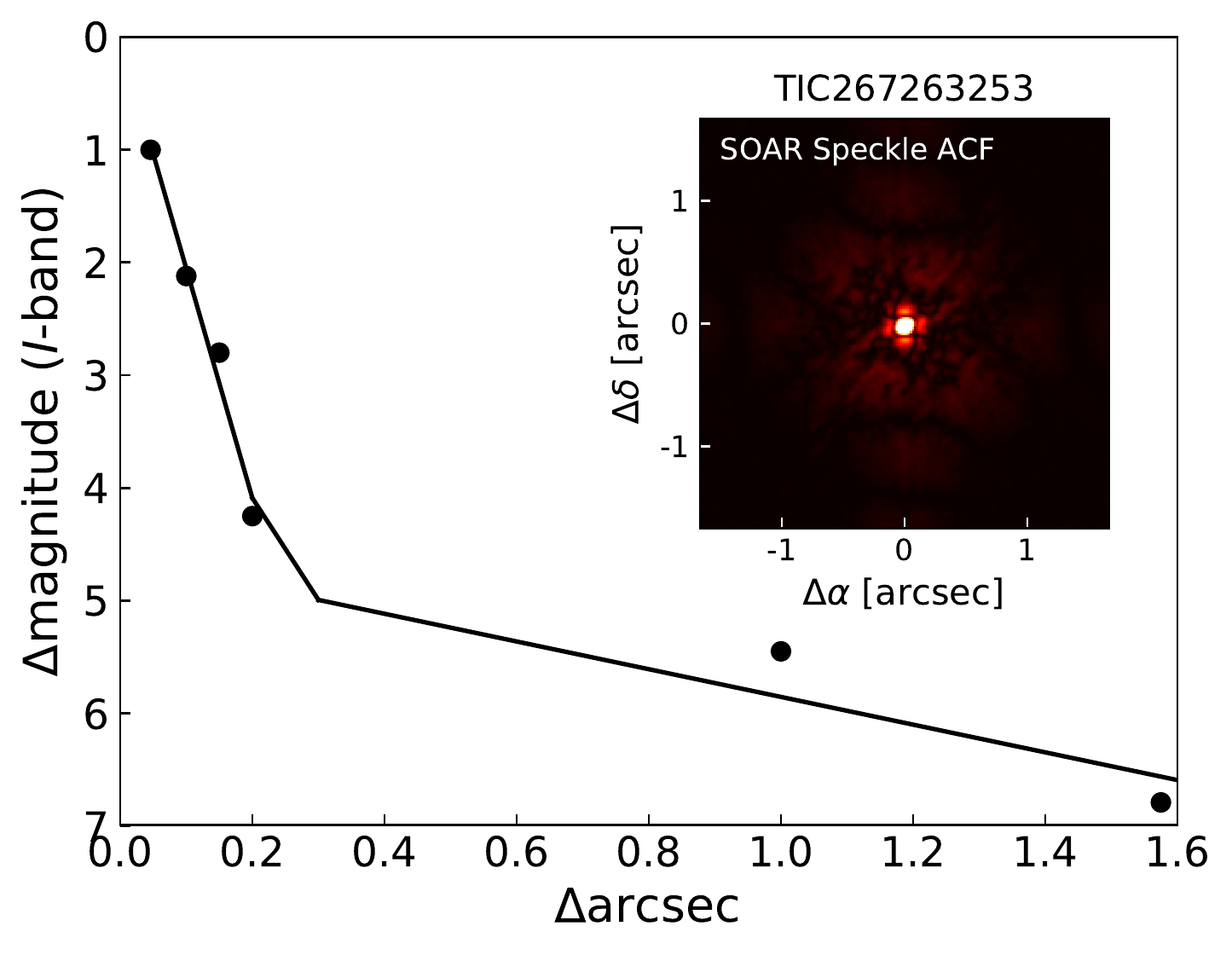}
\caption{Speckle auto-correlation function obtained in the
$I$-band, at the SOAR telescope. The black dots correspond to the 
5-$\sigma$ contrast curve for \stnameA\,\label{fig:speckle}. The solid line corresponds to the linear fit of the data, at separations smaller and larger than $\sim$ 0.2 arcsec.}
\end{figure}

The relatively large 21\arcsec pixels of TESS can result in photometric
contamination from nearby sources. These must be accounted for to rule out
astrophysical false positives, such as background eclipsing binaries, and
to correct the estimated planetary radius, initially derived from the
diluted transit in a blended light curve. We searched for close companions
to \stnameA\, with speckle imaging on the 4.1-m Southern Astrophysical
Research (SOAR) telescope \citep{tokovinin2018} on 2018 September 25 UT. The
5$\sigma$ detection sensitivity and auto-correlation function of the
observation are shown in Figure  \ref{fig:speckle}. 
We detected no nearby stars down to a magnitude difference of 4 within 0.2 \arcsec and 5 magnitudes within 1\,\arcsec\, of \stnameA\, in the SOAR images.

\subsection{CHIRON Radial Velocities}

As part of a two-week spectroscopic confirmation campaign of a handful of short-period transiting candidate planets detected by TESS, we obtained a total of 11 spectra of \stnameA\, using CHIRON (\citealp{chiron}), a fibre-fed high-resolution spectrograph, mounted on the SMARTS 1.5m telescope, at the Cerro Tololo Inter-American Observatory, in Chile.
The spectra were taken using a 2.7 arcsec diameter fibre on sky, with the high efficient image slicer mode. The exposure time was 1200\,seconds,  leading to a mean signal-to-noise-ratio (SNR) per pixel at 5500\,\AA\, of $\sim$\,45. From the spectra, we computed relative RV measurements using the cross-correlation technique, 
and correcting the night drift using Th-Ar spectra taken before and after each science exposure, as explained in \citet{wang2018}. This method is applied individually to a total of 41 orders covering the wavelength range between $\sim$ 4500\,-\,7000\,\AA\,. The mean RV uncertainty per epoch is $\sim$ 30 \ms, explained mainly due to the fast stellar rotation.
The resulting RVs are plotted in Figure \ref{fig:RV_curve1} and are also listed in Table \ref{tab:rv1}. The resulting CHIRON RVs of \stnameA\, revealed an amplitude of $\gtrsim$ 100 \ms, in phase with the orbital period detected in the TESS photometry. 
Additionally, from the cross-correlation-function (CCF) we computed the bisector velocity span (BVS), which are also listed in Table \ref{tab:rv1}. Figure \ref{fig:BVS_RV} shows the BVS versus the RVs. While the BVS scatter is large, they do not show a correlation with the RVs. In fact, the Pearson correlation coefficient between the CHIRON BVS and RVs is -0.14. 

\subsection{FEROS Radial Velocities}

\begin{figure}[h]
\vspace{0cm}\hspace{0cm} 
\includegraphics[scale=0.32,angle=0]{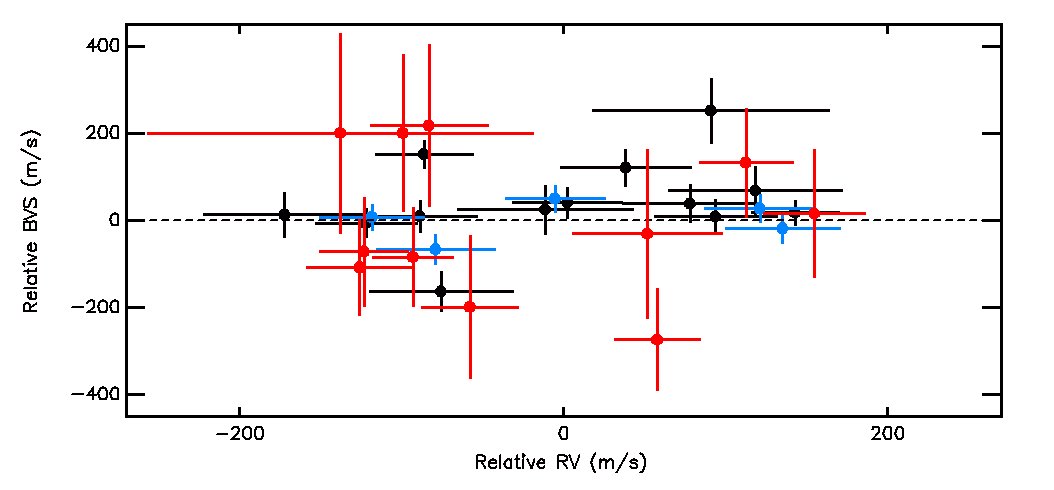}
\caption{Bisector velocity span as functions of the CORALIE (black dots), CHIRON (red dots) and FEROS (blue dots) radial velocities. An arbitrary offset is applied to the BVS values.
\label{fig:BVS_RV}}
\end{figure}

The CHIRON data were complemented with the Fiber-fed Extended Range Optical Spectrograph (FEROS; \citealt{feros}) spectra, using the simultaneous calibration technique (\citealp{baranne1996}). In this mode, the 2.0 \arcsec\,diameter science fibre is illuminated by the star, while the second fibre is illuminated by a Th-Ar reference lamp, which is used to correct for the RV drift.
We obtained five spectra of \stnameA\, with an exposure time of 300 seconds, leading to a mean SNR per resolution element of $\approx$100. The data reduction and RV calculation were performed with the \ceres\, pipeline \citep{ceres_code}. The reduction process is relatively standard for echelle spectra, and consists of a bias correction, flat field order tracing and optimal extraction. This method is applied to a total of 25 orders covering the wavelength region between $\sim$ 3900\,-\,6800\AA. Finally, each order is wavelength calibrated. The wavelength solution is accurate at the $\sim$ 2 \ms level, and is obtained from a total $\sim$ 1000 Th-Ar emission lines.
The RVs are computed via cross-correlation between each individual order with a binary numerical mask, and are corrected for the night drift using the Th-Ar spectra. The resulting velocities are plotted in Figure \ref{fig:RV_curve1}, and are listed in Table \ref{tab:rv1}. Similarly we computed FEROS BVS values, which are also listed in Table \ref{tab:rv1}, and plotted in Figure \ref{fig:BVS_RV}. As can be seen there is no obvious correlation between these two quantities. The Pearson correlation coefficient between the FEROS BVS and the corresponding RVs is 0.24.

\subsection{CORALIE Radial Velocities}

\begin{figure*}[h]
\vspace{0cm}\hspace{0cm} 
\includegraphics[scale=0.75]{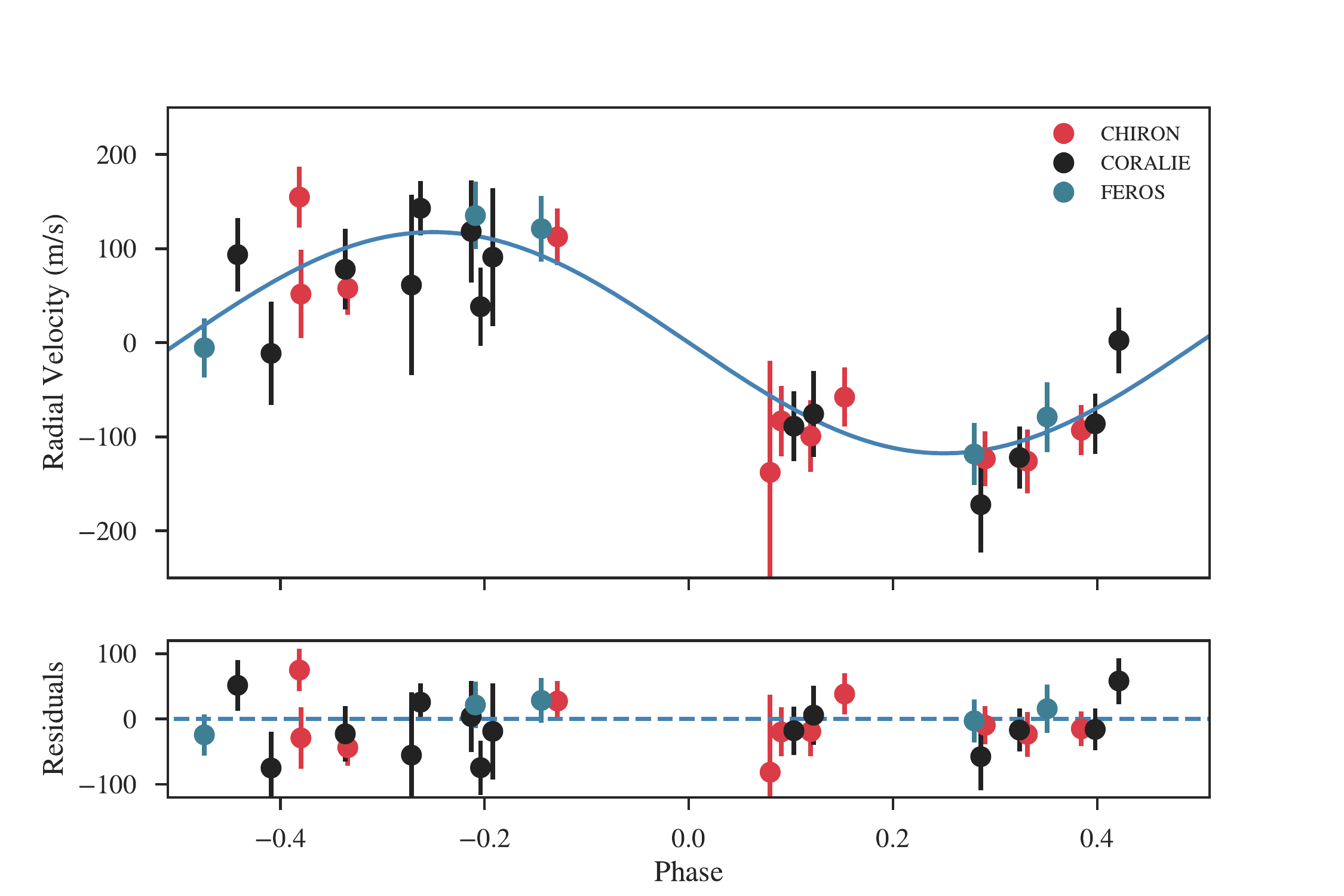}
\caption{Phase-folded RV curve of \stnameA. The black, red and blue points correspond to CORALIE, CHIRON and FEROS velocities, respectively. The best keplerian fit from the joint analysis is overplotted. The post-fit residuals are shown in the lower panel.
\label{fig:RV_curve1}}
\end{figure*}

Additional 14 spectra were obtained with the CORALIE high resolution spectrograph on the Swiss 1.2 m Euler telescope at La Silla Observatories, Chile \citep{coralie}, between 21 September and 28 October 2018. CORALIE is fed by a 2’’ science fibre and a secondary fibre with simultaneous Fabry-Perot for wavelength calibration. RVs were computed by cross-correlation with a binary G2 mask. Our exposures varied between 900 and 1800 seconds depending on the weather conditions. We obtain a final accuracy of 30-40 m/s which is limited by the stellar rotation due to broadening of the CCF. The resulting velocities are plotted in Figure \ref{fig:RV_curve1}, and are listed in Table \ref{tab:rv1}. Additionally, we computed the BVS values at each observing epoch, which are also listed in Table \ref{tab:rv1} and plotted in Figure \ref{fig:RV_curve1}. The Pearson linear coefficient between the CORALIE RVs and BVS values is 0.01.

\begin{table*}[h!]
\centering
\caption{Relative radial velocities of \stnameA\ \label{tab:rv1}}
\begin{tabular}{lrrrrr}
\hline\hline
\vspace{-0.3cm} \\
~~~BJD        & RV      & $\sigma_{RV}$  & BVS   & $\sigma_{BVS}$  & Instrument\\
 -2400000     & \ms     & \ms            & k\ms   & k\ms          &              \\
\hline \vspace{-0.3cm} \\
  58369.6043 & 186.0    &  30.8 & 0.1851  & 0.1446 &  CHIRON  \\
  58369.8027 &   89.0   &  26.0 & -0.1046 & 0.1149 &  CHIRON \\
  58370.6479 &  143.7   &  28.4 & 0.3017  & 0.1231 &  CHIRON \\
  58371.6716 &  -68.0   &  36.6 & 0.3703  & 0.1771 &  CHIRON \\
  58371.8098 &  -26.6   &  29.5 & -0.0291 & 0.1627 &  CHIRON \\
  58372.7658 &  -61.8   &  24.4 & 0.0857  & 0.1114 &  CHIRON \\
  58373.7375 &   82.9   &  45.8 & 0.1388  & 0.1910 &  CHIRON \\
  58375.6807 &  -51.9   &  35.9 & 0.3874  & 0.1831 &  CHIRON \\
  58379.7618 & -106.6   & 118.4 & 0.3702  & 0.2271 &  CHIRON \\
  58380.8013 &  -94.7   &  32.2 & 0.0617  & 0.1077 &  CHIRON \\
  58384.7577 &  -92.1   &  27.0 & 0.0977  & 0.1233 &  CHIRON \\
  58378.8372 & 2292.0   &  33.5 & -0.0310 & 0.0300  &  FEROS \\
  58380.8811 & 2091.9   &  36.0 & -0.1250 & 0.0320  &  FEROS \\
  58382.6972 & 2306.1   &  34.8 & -0.0770 & 0.0310  &  FEROS \\
  58384.7130 & 2052.5   &  31.6 & -0.0510 & 0.0280  &  FEROS \\
  58385.7289 & 2165.5   &  30.2 & -0.0090 & 0.0280  &  FEROS \\
  58382.6797 & 2178.9  &  53.2 & -0.1756 & 0.0532  & CORALIE \\
  58384.7383 & 1888.4  &  49.8 & -0.2310 & 0.0497  & CORALIE \\
  58390.6914 & 2121.9  &  95.2 & -0.8211 & 0.0952  & CORALIE \\
  58397.6797 & 2062.9  &  33.2 & -0.2030 & 0.0331  & CORALIE \\
  58398.6797 & 2138.6  &  41.3 & -0.2055 & 0.0412  & CORALIE \\
  58401.7109 & 1974.4  &  29.6 & -0.0919 & 0.0295  & CORALIE \\
  58404.6992 & 1984.9  &  43.9 & -0.4070 & 0.0439  & CORALIE \\
  58406.6328 & 2049.2  &  53.8 & -0.2191 & 0.0538  & CORALIE \\
  58407.5273 & 2151.5  &  72.5 &  0.0080 & 0.0724  & CORALIE \\
  58408.7462 & 1971.8  &  35.1 & -0.2345 & 0.0351  & CORALIE \\
  58409.6563 & 1938.5  &  30.8 & -0.2504 & 0.0307  & CORALIE \\
  58410.6250 & 2154.2  &  37.3 & -0.2356 & 0.0372  & CORALIE \\
  58411.6055 & 2098.9  &  40.0 & -0.1233 & 0.0400  & CORALIE \\
  58419.6172 & 2203.5  &  26.4 & -0.2266 & 0.0263  & CORALIE \\
\vspace{-0.3cm} \\\hline\hline
\end{tabular}
\end{table*}

\section{Host stars properties.}
\label{sec:host}

We first used the co-added FEROS spectra to estimate the
atmospheric parameters (\teff, \logg, \feh) and the projected rotational
velocity (\vsini) of \stnameA\,. Specifically, we use the \zaspe\ code
\citep{zaspe_code} to compare the observed spectra to a grid of synthetic models generated from the ATLAS9 model atmospheres \citep{castelli2004}. The comparison is performed in an
iterative fashion, only in the spectral regions that are more sensitive to
changes in the atmospheric parameters. These regions are computed automatically
from the gradient of the synthetic grid with respect to the atmospheric parameters.
Reliable uncertainties and correlations between the parameters are obtained
through Monte Carlo simulations that take into account systematic missmatches
between the optimal model and the observed spectrum. Then, we computed the stellar physical parameters, following  \citet{brahm2018}. This is done by combining the public broad band photometry, the {\it Gaia} DR2 parallaxes (\citealt{gaiadr2}), the derived atmospheric parameters, and the Yonsei-Yale isochrones (\citealt{yonseiyale}). For this, we use the Gaia parallax and a \texttt{BT-Settl-CIFIST} spectral energy distribution model having parameters equal to those derived through our spectroscopic analysis to
generate synthetic magnitudes. The set of synthetic magnitudes depend on the stellar
radius (\rstar) and the extinction coefficient (A$_V$). We then search for the distribution
of \rstar and A$_V$ by comparing the synthetic magnitudes to the observed ones
presented in Table \ref{tab:atm_par}. The exploration of the parameter space is performed with
the \texttt{emcee} package (\citealt{emcee_package}). We found that \stnameA\, is not significantly affected by
interstellar reddening. Finally, to obtain the mass, age and luminosity of the star, we compared its derived effective temperature and radius to that predicted by the Yonsei-Yale isochrones for different
masses and evolutionary stages, by fixing the metallicity of the isochrones to
the value obtained in the spectroscopic analysis. Again, the parameter
space was explored using the \texttt{emcee} package.
This procedure, through the estimation of the stellar mass and radius,
allows us to estimate a more precise value of \logg\, than the one obtained
with the spectroscopic analysis. Therefore, we performed a new \zaspe\ run fixing
\logg\ to the value obtained from \rstar\ and \mstar, and then we also repeated the
determination of these physical parameters. Using this method we obtained the following atmospheric parameters for \stnameA: \teff\,=\,6801 $\pm$ 56, \logg\,=\,4.22 $\pm$ 0.01 cm\,s$^{-2}$, \feh\,=\,+0.02 $\pm$ 0.03 dex and $v$\,sin$i$ = 15.4 $\pm$ 0.2 k\ms. 
Similarly, we obtained the following physical parameters for \stnameA\,: \mstar\,=\,1.44 $\pm$ 0.02 \msun, \rstar\,=\,1.57 $\pm$ 0.01 \rsun, L$_\star$ = 4.70$^{+0.18}_{-0.15}$ L$_\odot$ and Age = $1.30 \pm 0.14$ Gyr. 
These values are also listed in Table \ref{tab:atm_par}, and are those adopted for the rest of the paper. \newline \indent
For comparison, we also derived the stellar atmospheric parameters following the method presented in \citet{jones2011} and \citet{wang2018}. Briefly, we used ARES\,v2 \citep{Sousa2015} to measure the equivalent width (EW) of $\sim$ 150 Fe\,{\sc i} and Fe\,{\sc ii} absorption lines, from the co-added CHIRON template. These resulting EWs are then compared to the synthetic EWs generated by the \moog\ code \citep{Sneden1973}, that solves the radiative transfer equations under the assumptions of local thermodynamic equilibrium, using the \citet{kurucz1993} stellar atmosphere models. From this analysis we obtained the following atmospheric parameters for  HD\,2685: \teff = 6800 $\pm$ 70 K, \logg =  4.15 $\pm$ 0.15 and [Fe/H] = -0.08  $\pm$ 0.10, which are in good agreement with those derived by \zaspe. \newline \indent 
Finally, we also performed an analysis of the broadband spectral energy distribution (SED) of HD~2685 together with the {\it Gaia\/} DR2 parallax in order to determine an empirical measurement of the stellar radius, following \citet{Stassun:2018}. 
For this, we use the $B_T V_T$ magnitudes from {\it Tycho-2}, the Str\"{o}mgren $ubvy$ magnitudes from \citet{Paunzen:2015}, the $BVgri$ magnitudes from APASS, the $JHK_S$ magnitudes from {\it 2MASS}, the W1--W4 magnitudes from {\it WISE}, and the $G$ magnitude from {\it Gaia}. 
We performed a fit using Kurucz stellar atmosphere models, with the free parameters being the effective temperature, the surface gravity, the metallicity and the extinction, which we restricted to the maximum line-of-sight value from the dust maps of \citet{Schlegel:1998}.  The best fit parameters are $T_{\rm eff} = 6800 \pm 100$~K, $\log g = 4.0 \pm 0.25$, ${\rm [Fe/H]} = 0.00 \pm 0.25$, and $A_V = 0.09 \pm 0.02$. These values are consistent with those determined from the spectroscopic analysis. 
Finally, by taking the stellar bolometric flux computed from the unreddened model SED, and $T_{\rm eff}$ together with the {\it Gaia\/} DR2 parallax, gives a stellar radius of $R = 1.57 \pm 0.03$~R$_\odot$, in excellent agreement with that determined from the \zaspe\, code.

\begin{table}
\centering
\caption{Stellar parameters. \label{tab:atm_par}}
\begin{tabular}{lrr}
\hline\hline
Parameter               &         \stnameA         & Method/Source \\
\hline
\vspace{-0.3cm} \\
\teff (K)               &  6801 $\pm$ 56          &  \zaspe        \\
\logg \,(cm\,s$^{-2}$)  &  4.22 $\pm$ 0.01        &  \zaspe\,+\,\gaia    \\
{\rm [Fe/H]} (dex)      & +0.02 $\pm$ 0.03        &  \zaspe         \\
$v$\,sin$i$ (k\ms)        &  15.4 $\pm$ 0.2         &  \zaspe         \\

\mstar (\msun)          &  1.44 $\pm$ 0.02        &  \zaspe\,+\,\gaia\,+YY \\
\rstar (\rsun)          &  1.57 $\pm$ 0.01        &  \zaspe\,+\,\gaia\, \\
Distance (pc)       &  197.98 $^{+0.85}_{-0.70}$                &   \gaia\, \\ \vspace{-0.2cm} \\
L$_\star$ (L$_\odot$)   &  4.70$^{+0.18}_{-0.15}$  & \zaspe\,+\gaia\,+YY \\ \vspace{-0.2cm} \\
Age (Gyr)               &  1.3  $\pm$ 0.1         &  \zaspe+\gaia\,+YY \\ 
B (mag)                 & 10.05 $\pm$ 0.03        &  Tycho-2 \\
V (mag)                 &  9.59 $\pm$ 0.02        &  Tycho-2 \\
G (mag)                 & 9.5203 $\pm$ 0.0003     &  \gaia   \\
J (mag)                 & 8.825 $\pm$ 0.026       &  2MASS  \\
H (mag)                 & 8.651 $\pm$ 0.051       &  2MASS  \\ 
K (mag)                 & 8.595 $\pm$ 0.019       &  2MASS  \\           
\vspace{-0.3cm} \\\hline\hline
\end{tabular}
\end{table}

\section{Global analysis 
\label{sec:joint_fit}}

We performed a global analysis using both the photometric and spectroscopic data available. To do this, we used the EXOplanet traNsits and rAdIaL veLocity fittER (\exonailer\,) which is described in details in \citet{exonailer_code}, and available at GitHub.\footnote{https://github.com/nespinoza/exonailer}.
Briefly, we model the detrended light curve using the \batman\, code \citep{batman}, fitting simultaneously the limb-darkening coefficients with the rest of the transit parameters, and following the quadratic limb-darkening law presented in \citet{espinoza2016}. Similarly, the RV measurements were modelled using the \radvel\, package (\cite{radvel_code}). For the RVs, we included a jitter and RV offset as a free-parameter for each RV dataset. We first performed the fit with the eccentricity as a free parameter. We obtained a value of $e$ = 0.035$^{+0.080}_{-0.010}$, which is consistent with a circular orbit. Therefore, we repeated the analysis adopting a zero eccentricity.
Table \ref{glo_mod} lists 
the resulting transit values and the derived planetary parameters. Finally, from the derived stellar and planetary parameters we computed the equilibrium temperature (T$_{eq}$) for \plnameA\,. By adopting a zero albedo and assuming a tidally locked planet with no heat distribution ($\beta$ = 0.5;  \citealt{kaltenegger2011}) we obtained T$_{eq}$ = 2061 $\pm$ 21 K, for \plnameA.

\subsection{Searching for secondary eclipse and orbital variations}

We have examined the phase folded light curve for any variability outside of the transit, including orbital phase curves \citep{shporer2017} and secondary eclipse. We did not detect any sinusoidal variation along the orbital phase, which is expected given the system parameters and the sensitivity of the data. We attempted to measure the secondary eclipse, assuming its duration is identical to that of the transit and taking place exactly half an orbital period away from it. We measured a depth of 17$\pm$23 ppm,  meaning we do not detect the secondary eclipse and are able to place a $3 \sigma$ upper limit of 69 ppm on its depth (or a $2 \sigma$ upper limit of 46 ppm). This translates to a $3 \sigma$ upper limit on the geometric albedo in the TESS band of $A_g < 0.47$ (or a $2 \sigma$ upper limit of 0.31). Such an upper limit is consistent with the low geometric albedos found for hot Jupiter exoplanets in visible light \citep[e.g.,][]{heng2013, shporer2014, esteves2015, angerhausen2015}. In the above we have assumed that the thermal emission has a small to negligible contribution to the secondary eclipse, which in this system reaches only 12 ppm in the extreme case of no heat circulation between the day to night planet hemispheres. 

\begin{table}
\centering
\caption{Parameters obtained from the global modeling. \label{glo_mod}}
\begin{tabular}{lr}
\hline\hline
Parameter               &        Value  \\
\hline
\vspace{-0.2cm} \\
Light-curve parameters  &  \\\vspace{-0.2cm} \\
\ \  P (days)        & 4.12692$^{+0.00004}_{-0.00004}$   \\ \vspace{-0.2cm} \\
\ \  T$_c$ (BJD)     & 2458325.78290$^{+0.00017}_{-0.00015}$ \\ \vspace{-0.2cm} \\
\ \  $a$/\rstar      & 7.6996$^{+0.0278}_{-0.0336}$       \\ \vspace{-0.2cm}   \\
\ \  R$_P$ / \rstar  & 0.09481$^{+0.00028}_{-0.00026}$   \\ \vspace{-0.2cm} \\
\ \  $i$ (deg)       & 89.400 $^{+0.322}_{-0.334}$          \\ \vspace{-0.2cm} \\
\ \  q$_1$ (LCO)   & 0.161$^{+0.117}_{-0.056}$            \\ \vspace{-0.2cm} \\
\ \  q$_2$ (LCO)   & 0.585$^{+0.258}_{-0.279}$            \\\vspace{-0.2cm} \\
\ \  q$_1$ (CHAT)    & 0.071$^{+0.126}_{-0.051}$            \\ \vspace{-0.2cm} \\
\ \  q$_2$ (CHAT)    & 0.286$^{+0.345}_{-0.209}$            \\\vspace{-0.2cm} \\
\ \  q$_1$ (TESS)    & 0.146$^{+0.039}_{-0.031}$            \\ \vspace{-0.2cm} \\
\ \  q$_2$ (TESS)    & 0.389$^{+0.098}_{-0.089}$            \\\vspace{-0.2cm} \\
 & \\
RV parameters  &  \\\vspace{-0.2cm} \\
\ \  $e$                                  & 0 (fixed) \\ 
\ \  $K$ (m s$^{-1}$)                     & 117.6$^{+9.9}_{-8.7}$   \\
\ \  $\gamma_{\rm chiron}$ (km s$^{-1}$)  & 0.031 $\pm$ 0.012  \\
\ \  $\gamma_{\rm feros}$ (km s$^{-1}$)   & 2.170 $\pm$ 0.016  \\
\ \  $\gamma_{\rm coralie}$ (km s$^{-1}$)   & 2.061 $\pm$ 0.013  \\
\ \  $\sigma_{\rm chiron}$  (m s$^{-1}$)  & 11$^{+13}_{-8}$    \\
\ \  $\sigma_{\rm feros}$  (m s$^{-1}$)   & 9$^{+13}_{-7}$    \\
\ \  $\sigma_{\rm coralie}$  (m s$^{-1}$)   & 12$^{+18}_{-9}$    \\
 & \\
Planetary parameters  &  \\\vspace{-0.2cm} \\
\ \ $M_P$ ($M_J$) & 1.18 $\pm$ 0.09\\
\ \ $R_P$ ($R_J$) & 1.44 $\pm$ 0.01\\
\ \ $a$ (AU)      & 0.0568$^{+0.0003}_{-0.0002}$\\
\ \ T$_{eq}$ (K)  & 2061 $\pm$ 21 \\
\vspace{-0.3cm} \\\hline\hline
\end{tabular}
\end{table}

\section{Discussion}

\begin{figure}
\vspace{0cm}\hspace{0cm}
\includegraphics[scale=0.30,angle=0]{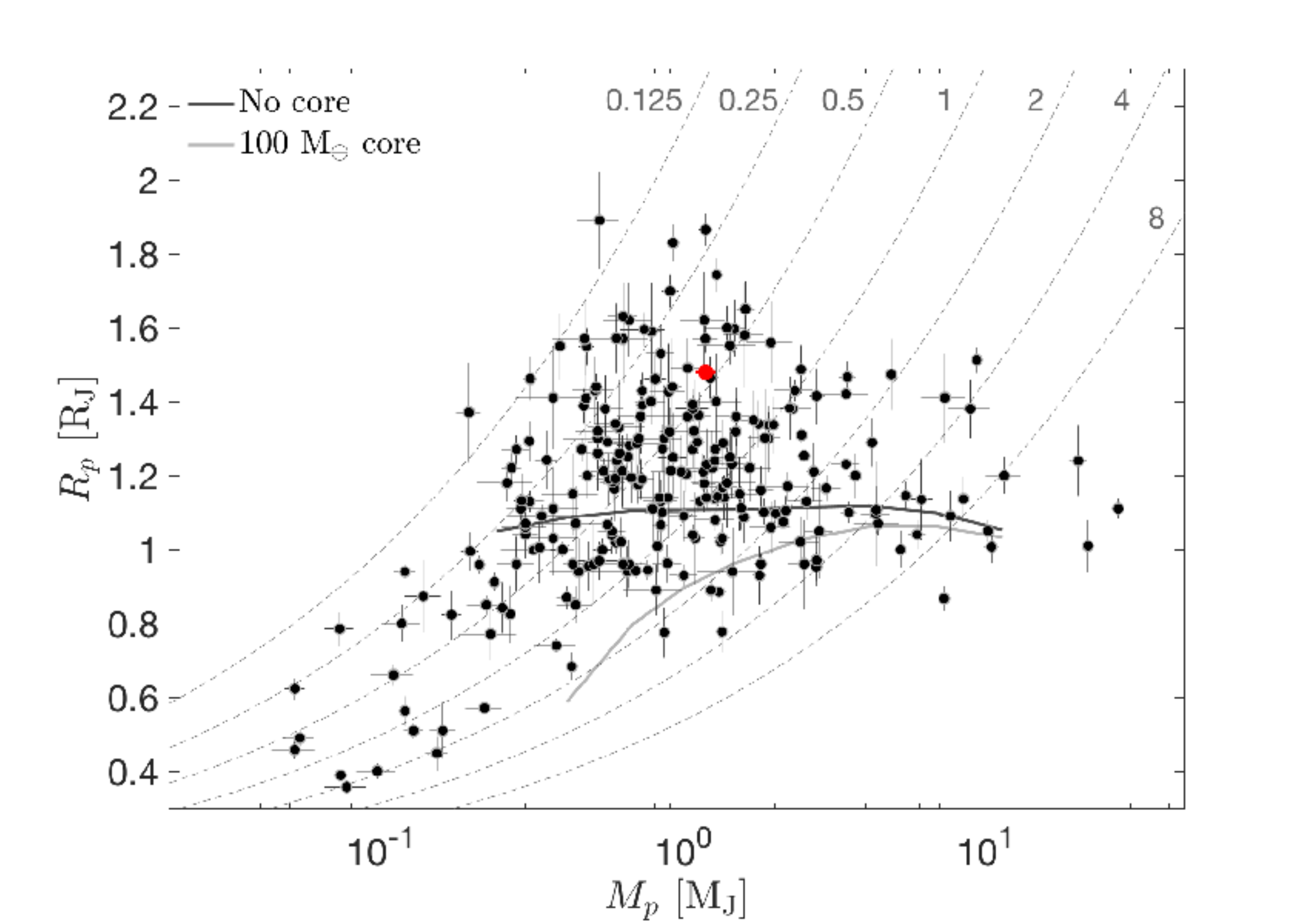}
\caption{Planet radius-mass diagram for transiting gas giant planets. The mass axis is plotted in logarithmic scale. The solid lines mark theoretical models taken from \citet{baraffe2014} for no solid core (black) and a massive core of 100~M$_{\rm Earth}$ (gray). The dashed lines mark isodensity contours, with the density values labeled at the top-right in gr cm$^{-3}$. The position of \plnameA\ shows it is an inflated gas giant planet, like other planets of similar mass. The plot includes planets whose radius uncertainty is smaller than 0.15 $R_J$ and whose mass uncertainty is smaller than a quarter of the measured mass. Data presented in this plot were obtained from the NASA Exoplanet Archive \citep{akeson2013} on October 29, 2018.
\label{fig:rad_mass}}
\end{figure}

\begin{figure}
\vspace{0cm}\hspace{0cm}
\includegraphics[scale=0.30,angle=0]{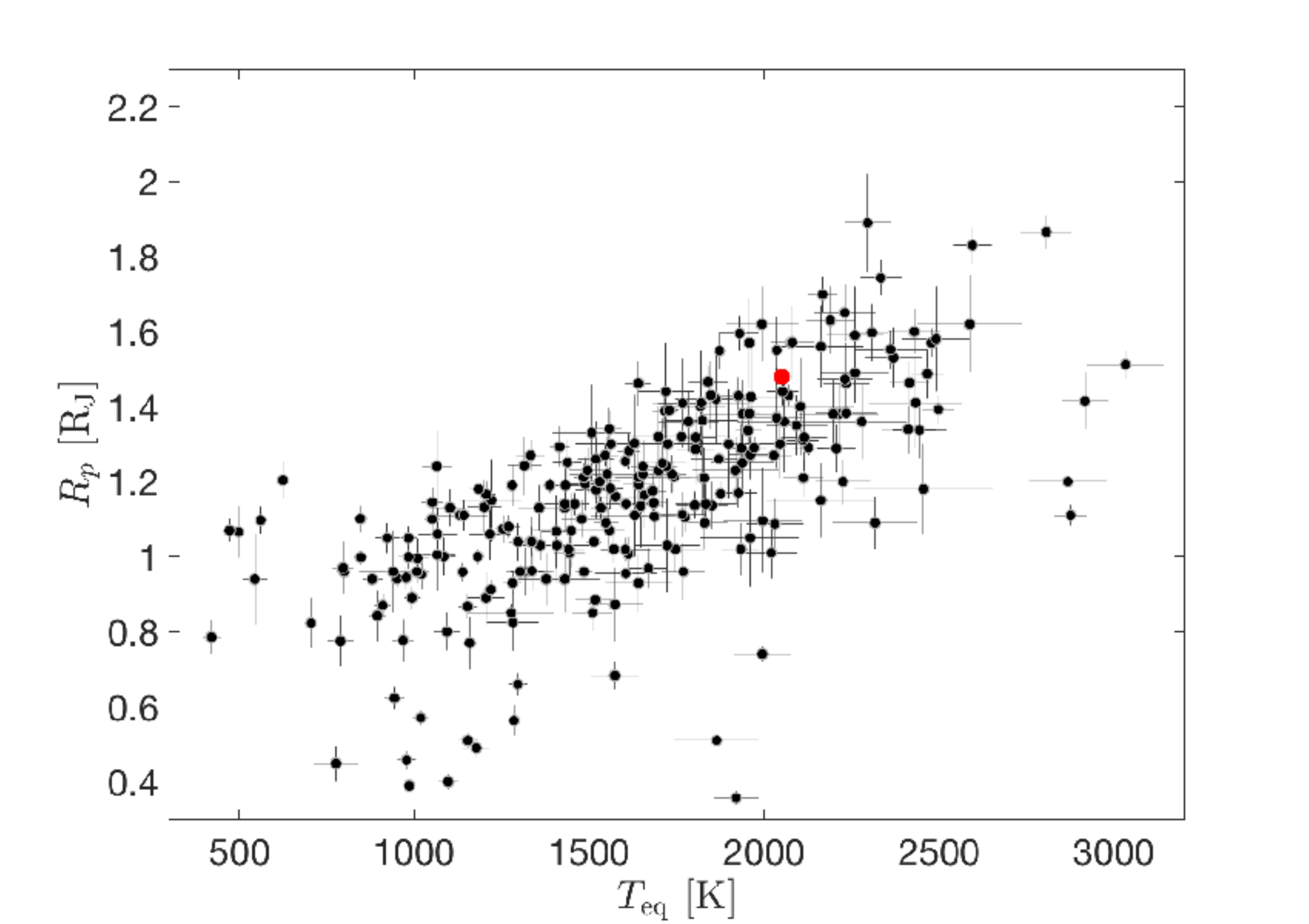}
\caption{Planet radius versus equilibrium temperature, showing the well known correlation between the two parameters. The position of \plnameA, marked in red, is in good agreement with this correlation. The plot includes planets whose radius uncertainty is smaller than 0.15 $R_J$ and whose mass uncertainty is smaller than a quarter of the measured mass.
Data presented in this plot were obtained from the NASA Exoplanet Archive \citep{akeson2013} on October 29, 2018.
\label{fig:rad_teq}}
\end{figure}

\begin{figure}
\vspace{0cm}\hspace{0cm} 
\includegraphics[scale=0.30,angle=0]{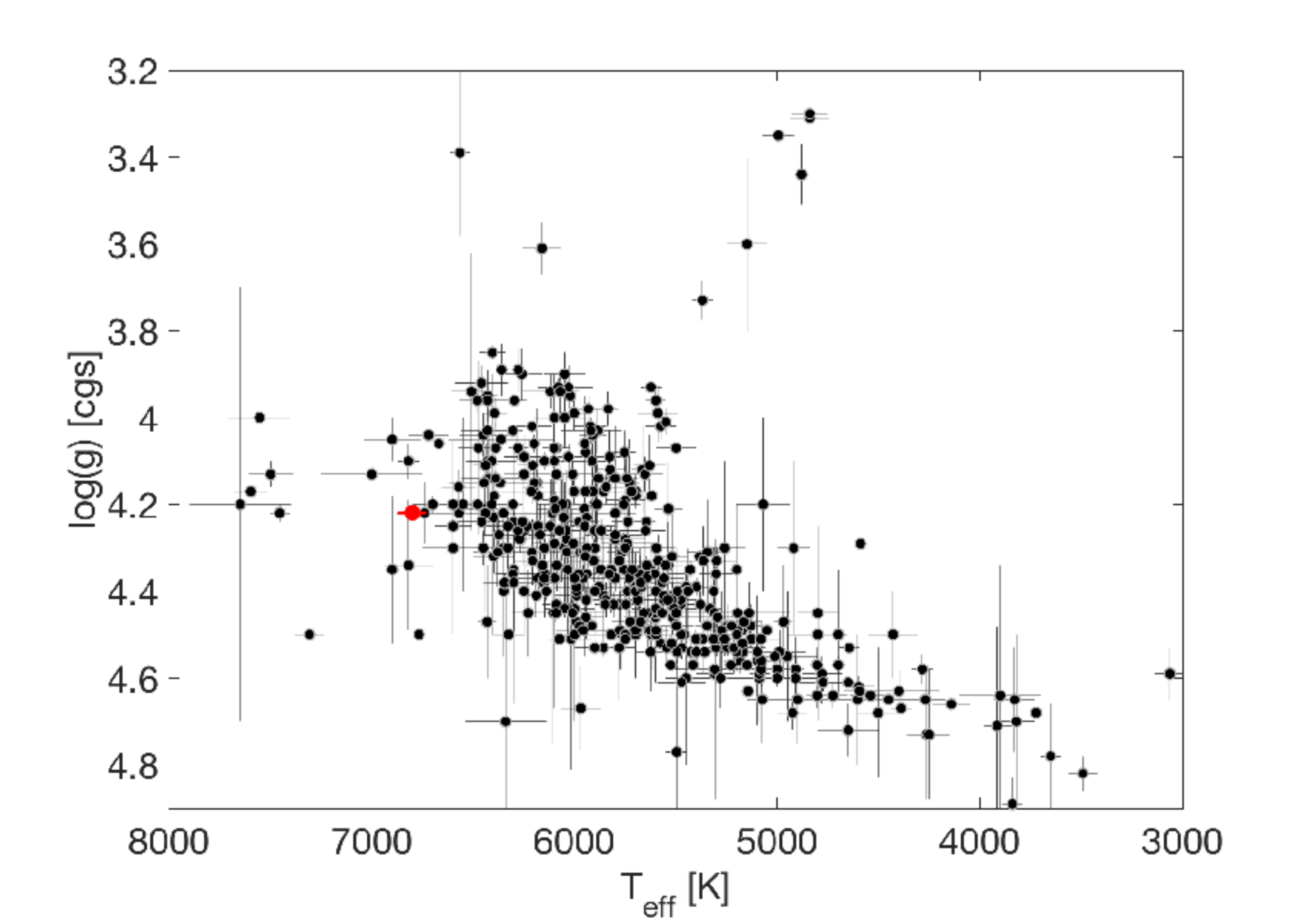}
\caption{Stellar surface gravity (in cgs units) versus effective temperature of all known transiting planet host stars where the planet's mass and radius are both measured. The red dot shows the position of \stnameA. Data presented in this plot were obtained from the NASA Exoplanet Archive \citep{akeson2013} on October 29, 2018.
\label{fig:logg_teff}}
\end{figure}

In this paper we have presented the radial velocity confirmation of \plnameA, a transiting {\it Hot-Jupiter} observed in Sector 1 of the TESS mission. We obtained CHIRON, CORALIE and FEROS spectroscopic data, from which we confirmed the planetary nature of the transit signal detected by TESS. 
In addition, we obtained CHAT and LCOGT photometric data during one  transit event. From the joint analysis we derived the following parameters for \plnameA\,: $P$=4.12692$\pm$0.00004 days, $M_p$=1.18 $\pm$ 0.09 $M_J$ and $R_p$=1.44 $\pm$ 0.01 $R_J$. We also measured a very low eccentricity ($e$ = 0.035$^{+0.08}_{-0.01}$), which is consistent with a circular orbit.
\newline \indent
Figure \ref{fig:rad_mass} shows the planetary radius versus mass for known transiting gas giants. The red dot corresponds to \plnameA. Different iso-density contours are over-plotted (dashed grey lines). Also for comparison, the solid lines mark theoretical models from \citet{baraffe2014}, for no solid core and 100 M$_\oplus$ core planets. 
Similarly, Figure \ref{fig:rad_teq} shows the planetary radius versus the planet's equilibrium Temperature. It can be clearly seen from these two plots that \plnameA\, is an inflated {\it Hot-Jupiter},
whose large radius cannot be explained by current theoretical models of planetary internal structure (e.g. \citealt{baraffe2014}) while it is likely explained by the large irradiation received by the parent star (e.g. \citealt{hartman2011}; \citealt{weiss2013}).  \newline \indent
Figure \ref{fig:logg_teff} shows the position of \stnameA\, in the stellar $\log(g)$ versus \teff parameter space for transiting planets host stars. It can be seen that \stnameA\, is among the hottest stars with a known transiting gas giant planet, with only about a dozen host stars at a similar temperature or hotter. \newline \indent


\subsection{Follow-up prospects}

%

\stnameA\, has a visual magnitude of $V$ = 9.6 mag, meaning it is among the brightest stars with a transiting giant planets, making it a valuable target for detailed follow-up characterization.

Apart from its high apparent brightness (V = 9.6 mag), \stnameA\, is a relatively hot (\teff = 6801 $\pm$ 56\,K) and fast rotating star ($v$\,sin$i$ = 15.4 $\pm$ 0.2 k\ms), making this object an excellent target for studying the projected spin-orbit angle via the Rossiter-McLaughlin effect. Knowing the stellar obliquity in transiting giant planets can give us important information about different planetary migration processes.
While some of these mechanism predict a damping in the spin-orbit primordial misalignment, if any (\cite{cresswell2007}), some others are expected to alter the mutual inclination angle (e.g. \citealt{fabrycky2007}). For this reason, knowing the obliquity angle can help us to distinguish between different migrations scenarios \citep{wang2018a}. 
Moreover, \citet{winn2010} showed that hotter stars (\teff $\gtrsim$ 6250 K) hosting {\it Hot-Jupiters} have larger obliquity compared to cooler stars. Based on this result, \plnameA\, might be expected to be in a high obliquity orbit. From the derived stellar and planet parameters, we predict an RV semi-amplitude during the transit of $\approx$92 \ms (assuming an aligned orbit), which can be easily detected given the RV precision attained for this object. Thus this target provides us with an excellent opportunity to further study this observational trend in the high \teff regime (see \citealt{addison2018} for an updated version of this correlation). \newline \indent
On the other hand, transiting exoplanets around bright stars provide a great laboratory for probing physical properties of these alien worlds. Arguably, the most intriguing and accessible of those characteristics is their atmospheric make-up. Studying exoplanetary atmospheres gives us clues about the formation and evolution history of the planetary system, the composition of the initial protoplanetary disk in which the planet was formed, the location of that formation (\citealt{madhusudhan2014}; \citealt{mordasini2016}), as well as the internal structure of the planet (\citealt{dorn2015}).
There are a variety of techniques through which minute signatures of exoplanetary atmospheres are detected. The most effective of these has been the transmission spectroscopy that searches for atmospheric imprints on the traversing stellar light (\citealt{seager2000}). This is mainly done through either performing mid-to-low resolution, highly time-resolved spectroscopy using multi-object spectrographs from the ground (e.g. FORS2; \cite{bean2011}, \cite{sedaghati2017}. IMACS; \citealt{jordan2013}, \cite{espinoza2018}) or space (e.g. HST and Spitzer; \citealt{sing2016}), or high resolution stable spectroscopy (e.g. HARPS; \citealt{allart2017}).
The significance of the detection of various atomic and/or molecular species heavily relies on  the brightness of the host star, and the scale height of the exo-atmosphere being probed. Consequently, \plnameA\, is an ideal candidate for atmospheric follow-up studies on both accounts. The 9.6 visual magnitude of the host star allows for obtaining high SNR spectra with relatively short exposure times, essential for performing transmission spectroscopy.  Additionally, assuming a H/He-dominated atmosphere, the extended scale height of the atmosphere ($ \textrm{H} = \frac{k_\textrm{B} T_{eq}}{\mu_m g} \approx 454$ km) due to the large equilibrium temperature expected, leads to significant atmospheric signals, well within the reach of current and future instrumentation (e.g. ESPRESSO, JWST).

\begin{acknowledgements}

S.W. and J.S.W. thank the Heising-Simons Foundation for their generous support.
CZ is supported by a Dunlap Fellowship at the Dunlap Institute for
Astronomy \& Astrophysics, funded through an endowment established by the
Dunlap family and the University of Toronto.
AV's work was performed under contract with the California Institute of Technology (Caltech)/Jet Propulsion Laboratory (JPL) funded by NASA through the Sagan Fellowship Program executed by the NASA Exoplanet Science Institute.
A.J.\ acknowledges support from FONDECYT project 1171208, CONICYT project Basal AFB-170002, and by the Ministry for the Economy, Development, and Tourism's Programa Iniciativa Cient\'{i}fica Milenio through grant IC\,120009, awarded to the Millennium Institute of Astrophysics (MAS). 
We acknowledge the use of TESS Alert data, which is currently in a beta test phase, from pipelines at the TESS Science Office and at the TESS Science Processing Operations Center.
This research has made use of the Exoplanet Follow-up Observation Program website, which is operated by the California Institute of Technology, under contract with the National Aeronautics and Space Administration under the Exoplanet Exploration Program.
This paper includes data collected by the TESS mission, which are publicly available from the Mikulski Archive for Space Telescopes (MAST).
This work makes use of observations from the LCOGT network.
\end{acknowledgements}




\begin{thebibliography}{10}


\bibitem[Addison et al.(2018)]{addison2018} Addison, B. C., Wang, S., Johnson, M. C., et al. 2018, AJ, 156, 197

\bibitem[Akeson et al.(2013)]{akeson2013} Akeson, R.~L., Chen, X., Ciardi, D., et al.\ 2013, \pasp, 125, 989 

\bibitem[Allart et al.(2017)]{allart2017} Allart, R:, Lovis, C., Pino, L., et al. 2017, A\&A, 606, 144

\bibitem[Angerhausen et al.(2015)]{angerhausen2015} Angerhausen, D., DeLarme, E., \& Morse, J.~A.\ 2015, \pasp, 127, 1113 


\bibitem[Baraffe et al.(2014)]{baraffe2014} Baraffe, I., Chabrier, G., Fortney, J., \& Sotin, C.\ 2014, Protostars and Planets VI, 763 

\bibitem[Baranne et al.(1996)]{baranne1996} Baranne, A., Queloz, D., Mayor, M. et al. 1996, \aap, 119, 373

\bibitem[Barclay et al.(2018)]{barclay2018} Barclay, T., Pepper, J. \& Quintana, E. 2018, ArXiv e-prints, arXiv:1804.05050

\bibitem[Bean et al.(2011)]{bean2011} Bean, J.~L., D{\'e}sert, J.-M., Kabath, P., et al. 2011, ApJ, 732, 92

\bibitem[Benneke \& Seager(2012)]{benneke2012} Benneke B.\& Seager S., 2012, ApJ, 753, 100

\bibitem[Borucki et al.(2010)]{kepler} Borucki, W. J., Koch, D., Basri, G., et al. 2010, Science, 327, 977

\bibitem[Brahm et al.(2017a)]{ceres_code} Brahm, R., Jord\'an, A. \& Espinoza, N. 2017a, PASP, 129, 34002

\bibitem[Brahm et al.(2017b)]{zaspe_code} Brahm R., Jord\'an A., Hartman J. \& Bakos G. 2017b, MNRAS, 467, 971

\bibitem[Brahm et al.(2018)]{brahm2018} Brahm, R., Espinoza, N., Jord\'an, A. et al. 2018, MNRAS 477, 2572

\bibitem[Broeg et al.(2013)]{cheops} Broeg, C., Fortier, A., Ehrenreich, D., et al. 2013, European Physical Journal Web of Conferences 47,  03005

\bibitem[Brown et al.(2013)]{brown2013} Brown, T.~M., Baliber, N., Bianco, F.~B., et al.\ 2013, \pasp, 125, 1031 


\bibitem[Castelli \& Kurucz(2004)]{castelli2004} Castelli, F., \& Kurucz, R: L. 2004, arXiv:0405087

\bibitem[Collins et al.(2017)]{collins2017} Collins, K.~A., Kielkopf, J.~F., Stassun, K.~G., \& Hessman, F.~V.\ 2017, \aj, 153, 77 

\bibitem[Cresswell et al.(2007)]{cresswell2007} Cresswell, P., Dirksen, G., Kley, W., \& Nelson, R. P. 2007, A\&A, 473, 329


\bibitem[Dorn et al.(2015)]{dorn2015} Dorn, C., Khan, A., Heng, K., et al. 2015, A\&A, 577, 83

\bibitem[Dressing \& Charbonneau(2013)]{dressing2013} Dressing, C.~D. \& Charbonneau, D. 2013, ApJ, 767, 95


\bibitem[Espinoza et al.(2016)]{exonailer_code} Espinoza, N., Brahm, R., Jord\'an, A., et al. 2016, ApJ, 830, 43

\bibitem[Espinoza \& Jord\'an(2016)]{espinoza2016} Espinoza, N. \& Jord\'an, A. 2016, MNRAS, 457, 3573

\bibitem[Espinoza et al.(2018)]{espinoza2018} Espinoza, N., Rackham, B.~V.,  Jord\'an, A., et al. 2018, submitted to MNRAS (arXiv:1807.10652)

\bibitem[Esteves et al.(2015)]{esteves2015} Esteves, L.~J., De Mooij, E.~J.~W., \& Jayawardhana, R.\ 2015, \apj, 804, 150 


\bibitem[Fabrycky \& Tremaine(2007)]{fabrycky2007} Fabrycky, D. \& Tremaine, S. 2007, ApJ, 669, 1298

\bibitem[Foreman-Mackey et al.(2013)]{emcee_package} Foreman-Mackey, D., Hogg, D.~W., Lang, D. \& Goodman, J. 2013, PASP, 125, 306

\bibitem[Foreman-Mackey et al.(2017)]{foreman-mackey2017} Foreman-Mackey, D., Agol, E., Ambikasaran, S. \& Angus, R. 2017, AJ, 154, 220

\bibitem[Fulton et al.(2018)]{radvel_code} Fulton B. J., Petigura E. A. \& Blunt S., Sinukoff E., 2018, PASP, 130, 986


\bibitem[Gaia collaboration et al.(2018)]{gaiadr2} Gaia collaoration, Brown, A.~G.~A., Vellenari, A., et al. 2018, ArXiv e-prints, arXiv:1804.09365

\bibitem[Gandolfi et al.(2018)]{gandolfi2018} Gandolfi, D., Barrag\'an, O., Livingston, J. H., et al. 2018, ArXiv e-prints, arXiv:1809.07573

\bibitem[Gardner et al.(2006)]{jwst} Gardner, J. P., Mather, J. C., Clampin, M., et al. 2006, Space Sci. Rev. 123, 485


\bibitem[Guillot \& Showman(2002)]{guillot2002} Guillot, T., \& Showman, A. P. 2002, A\&A, 385, 156


 \bibitem[Hartman et al.(2011)]{hartman2011} Hartman, J. D., Bakos, G. \'A., Torres, G., et al. 2011, ApJ, 742, 59
 
  \bibitem[Hartman et al.(2018)]{hartman:2018} Hartman, J. D., Bakos, G. \'A., Bayliss, D., et al. 2018, AJ submitted, arXiv:1809.01048
 
 \bibitem[Heng \& Demory(2013)]{heng2013} Heng, K., \& Demory, B.-O.\ 2013, \apj, 777, 100 

\bibitem[Huang et al.(2018)]{huang2018} Huang, C. X., Burt, J., Vanderburg, A., et al. 2018, ArXiv e-prints, arXiv:1809.05967




\bibitem[Kaltenegger \& Sasselov(2011)]{kaltenegger2011} Kaltenegger, L. \& Sasselov, D. 2011, ApJ, 736, L25

\bibitem[Kaufer et al.(1999)]{feros} Kaufer, A., Stahl, O., Tubbesing, S. et al. 1999, The Messenger 95, 8

\bibitem[Kreidberg(2015)]{batman} Kreidberg, L. 2015, PASP, 127, 1161

\bibitem[Kurucz(1993)]{kurucz1993} Kurucz, R.\ 1993, ATLAS9 Stellar Atmosphere Programs and 2 km/s grid.~Kurucz CD-ROM No.~13.~ Cambridge, Mass.: Smithsonian Astrophysical Observatory, 1993.


\bibitem[Jenkins et al.(2016)]{jenkins2016} Jenkins, J. M., Twicken, J. D., McCauliff, S., et al. 2016, in Proc. SPIE, Vol. 9913, Software and Cyberinfrastructure for Astronomy IV, 99133E

\bibitem[Jenkins(2017)]{jenkins2017} Jenkins, J.~M. (ed.)\ 2017, Kepler Data Processing Handbook, KSCI-19081-002

\bibitem[Jones et al.(2011)]{jones2011} Jones, M.~I., Jenkins, J.~S., Rojo, P., \& Melo, C.~H.~F.\ 2011, \aap, 536, A71 

\bibitem[Jones et al.(2017)]{jones2017} Jones, M.~I., Brahm, R:, Wittenmyer, R. A., et al. 2017, A\&A, 602, 58

\bibitem[Jord\'an et al.(2013)]{jordan2013} Jord\'an, A., Espinoza, N., Rabus, M., et al. 2013, ApJ, 778, 184

\bibitem[Jord\'an et al.(2018)]{jordan:2018} Jord\'an, A., Brahm, R:, Espinoza, N., et al. 2018, ArXiv e-prints, arXiv:1809.08879


\bibitem[Madhusudhan et al.(2014)]{madhusudhan2014} Madhusudhan, N., Amin, M.~A. \& Kennedy, G.~M. 2014, ApJ, 794, L12 

\bibitem[Matsumura et al.(2010)]{matsumura2010} Matsumura, S., Peale, S.~J. \& Rasio, F.~A. 2010, ApJ, 725, 1995 

\bibitem[Mordasini et al.(2016)]{mordasini2016} Mordasini, C., van Boekel, R:, Molliere, P., et al. 2016, ApJ, 832, 41

\bibitem[Mulders et al.(2015)]{mulders2015} Mulders, G. D., Pascucci, I. \& Apai, D. 2015, ApJ, 814, 130


\bibitem[Paunzen(2015)]{Paunzen:2015} Paunzen, E.\ 2015, \aap, 580, A23


\bibitem[Queloz et al.(2000)]{queloz2000} Queloz, D., Eggenberger, A., Mayor, M., et al. 2000, A\&A, 359, L13

\bibitem[Queloz et al.(2001)]{coralie} Queloz, D., Mayor, M., Udry, S., et al. 2001, The Messenger, 105, 1


\bibitem[Rasio et al.(1996)]{rasio1996} Rasio, F.~A., Tout, C.~A., Lubow, S.~H. \& Livio, M. 1996, ApJ, 470, 1187

\bibitem[Ricker et al.(2015)]{tess} Ricker, G. R., Winn, J. N., Vanderspek, R., et al. 2015, Journal of Astronomical Telescope, Instruments and Systems, 1, 014003


\bibitem[Schlegel et al.(1998)]{Schlegel:1998} Schlegel, D.~J., Finkbeiner, D.~P., \& Davis, M.\ 1998, \apj, 500, 525

\bibitem[Seager \& Sessalov(2000)]{seager2000} Seager, S., \& Sessalov, D.~D. 2000, ApJ, 537, 916.

\bibitem[Sedaghati et al.(2017)]{sedaghati2017} Sedaghati, E., Boffin, H.~M.~J., MacDonald, R.~J., et al. 2017, Nature, 549, 238

\bibitem[Shporer(2017)]{shporer2017} Shporer, A.\ 2017, \pasp, 129, 072001 

\bibitem[Shporer et al.(2014)]{shporer2014} Shporer, A., O'Rourke, J.~G., Knutson, H.~A., et al.\ 2014, \apj, 788, 92 

\bibitem[Sing et al.(2016)]{sing2016} Sing, D.~K., Fortney, J.~J., Nikolov, N., et al. 2016, Nature, 529, 59

\bibitem[Sneden(1973)]{Sneden1973} Sneden, C.\ 1973, \apj, 184, 839

\bibitem[Snellen et al.(2013)]{snellen2013} Snellen I. A. G., de Kok, R. J., le Poole R., Brogi M. \& Birkby J. 2013, ApJ, 764, 182

\bibitem[Sousa et al.(2015)]{Sousa2015} Sousa, S.~G., Santos, N.~C., Adibekyan, V., Delgado-Mena, E., \& Israelian, G.\ 2015, \aap, 577, A67 

\bibitem[Stassun et al.(2018)]{Stassun:2018} Stassun, K.~G., Corsaro, E., Pepper, J.~A., \& Gaudi, B.~S.\ 2018, \aj, 155, 22


\bibitem[Thorngren et al.(2016)]{thorngren2016} Thorngren, D.~P., Fortney, J.~J.; Murray-Clay, R.~A. \& Lopez, E.~D. 2016, ApJ, 831, 64

\bibitem[Tokovinin et al.(2013)]{chiron} Tokovinin, A., Fischer, D.~A., Bonati, M., et al.\ 2013, \pasp, 125, 1336

\bibitem[Tokovinin(2018)]{tokovinin2018} Tokovinin, A. 2018, PASP, 130, 35002

\bibitem[Wang et al.(2018a)]{wang2018a} Wang, S., Addison, B., Fischer, D.~A., et al.\ 2018, \aj, 155, 70  

\bibitem[Wang et al.(2018b)]{wang2018} Wang, S., Jones, M., Shporer, A., et al.\ 2018, ArXiv e-prints, arXiv:1810.02341 

\bibitem[Weiss et al.(2013)]{weiss2013} Weiss, L. M., Marcy, G. W., Rowe, J. F., et al. 2013, ApJ, 768, 14

\bibitem[Winn et al.(2010)]{winn2010} Winn, J. N., Fabrycky, D:,Albrecht, S., \& Johnson, J. A. 2010, ApJ, 718, 145L

\bibitem[Winn \& Fabrycky(2015)]{winn2015} Winn, J.~N. \& Fabrycky, D.~C. 2015, ARA\&A, 53, 409



\bibitem[Yi et al.(2001)]{yonseiyale} Yi, S., Demarque, P, Kim, Y.-C., et al. 2001, ApJS, 136, 417

\end{thebibliography}
\end{document}